\documentclass[prl,twocolumn]{revtex4}
\usepackage{amsmath}
\usepackage{amssymb}
\usepackage{amsfonts}

\usepackage{graphicx}
\usepackage{float}

\usepackage{amsthm}

\newcommand{\paren}[1]{{\left( #1 \right)}}

\renewcommand{\Im}{{\rm Im}}

\makeatletter
\def\loweq@align#1#2{\lower.6ex\vbox{\baselineskip\z@skip\lineskip\z@
    \ialign{$\m@th#1\hfil##\hfil$\crcr#2\crcr=\crcr}}}
\def\lowsim@align#1#2{\lower.6ex\vbox{\baselineskip\z@skip\lineskip\z@
    \ialign{$\m@th#1\hfil##\hfil$\crcr#2\crcr\sim\crcr}}}
\def\geqq{\mathrel{\mathpalette\loweq@align >}}
\def\leqq{\mathrel{\mathpalette\loweq@align <}}
\def\grsim{\mathrel{\mathpalette\lowsim@align >}}
\def\lesssim{\mathrel{\mathpalette\lowsim@align <}}
\def\gsim{\mathrel{\mathpalette\lowsim@align >}}
\def\lsim{\mathrel{\mathpalette\lowsim@align <}}
\makeatother
\newcommand{\grless} 
{ {\, \raise-.24em\hbox{$<$} \hspace{-0.8em} \raise.31em\hbox{$>$}\, } }
\newcommand{\lessgr} 
{ {\, \raise-.24em\hbox{$>$} \hspace{-0.8em} \raise.31em\hbox{$<$}\, } }
\newfont{\bg}{cmr10 scaled\magstep4}                    
\newcommand{\bigzerou}{\smash{\lower1.7ex\hbox{\bg 0}}}

\newcommand{\nn}{\nonumber \\ }

\newcommand{\di}{\displaystyle}

\newcommand{\crl}[1]{[-\infty,\infty]}

\newcommand{\ket}[1]{|{#1}\rangle}
\newcommand{\kb}[1]{|{#1}\rangle\!\langle{#1}|}

\newcommand{\bra}[1]{\langle{#1}|}
\newcommand{\bkt}[2]{\langle{#1}|{#2}\rangle}
\newcommand{\Ref}[1]{(\ref{#1})}

\newcommand{\av}[1]{\langle#1\rangle}

\newcommand{\f}{\frac}
\newcommand{\llim}{l_{\rm im}}
\newcommand{\llm}{l_{\rm \ell m}}
\newcommand{\lmd}{l_{\rm md}}
\newcommand{\lid}{l_{\rm id}}

\newcommand{\xl}{x_{\rm \ell}}

\newcommand{\xm}{x_{\rm m}}

\newcommand{\am}{a_{\rm m}}
\newcommand{\si}{s_{\rm i}}
\newcommand{\Phii}{\Phi_{i}}
\newcommand{\Phimi}{\Phi_{{\rm m},i}}
\newcommand{\Phili}{\Phi_{{\rm \ell},i}}

\newcommand{\Phif}{\Phi_{f}}
\newcommand{\psii}{i}
\newcommand{\psif}{f}
\renewcommand{\d}{{\rm d}}
\newcommand{\cw}{\circlearrowright}
\newcommand{\ccw}{\circlearrowleft}
\newcommand{\Psii}{\Psi_{i}}

\renewcommand{\a}{\alpha}
\renewcommand{\AA}{{\cal A}}
\newcommand{\degr}{\,{}^\circ}

\newcommand{\m}{{\rm m}}
\newcommand{\micm}{\mu{\rm m}}
\newcommand{\sect}[1]{\paragraph*{{#1}.---}}

\begin{document}

\title{On Amplification by Weak Measurement}
\author{Tatsuhiko Koike}
\email{koike@phys.keio.ac.jp}
\author{Saki Tanaka}
\email{stanaka@rk.phys.keio.ac.jp}
\affiliation{Department of Physics, Keio University, Yokohama 
  223-8522 Japan}

\date{August, 2011}

\begin{abstract}
  We analyze the amplification 
  by 
  the Aharonov-Albert-Vaidman 
  weak quantum measurement on 
  a Sagnac interferometer 
  [P.~B.~Dixon et al., 
  Phys. Rev. Lett. 102, 173601 (2009)] 
  up to all orders of the
  coupling strength between the measured system and the measuring
  device. 
  The 
  amplifier 
  transforms a small tilt of a mirror 
  into 
  a large transverse displacement 
  of the laser beam. 
  The conventional analysis has shown that 
  the measured value 
  is proportional to
  the weak value, so that 
  the amplification 
  can be made arbitrarily large 
  in the cost of decreasing output laser intensity. 
  It is shown that the measured displacement and the 
  amplification factor 
  are in fact not proportional to the weak value and 
  rather 
  vanish in the limit of infinitesimal output intensity. 
  We derive the optimal overlap of the pre- and
  post-selected states with which the amplification become maximum. 
  We also show that 
  the nonlinear effects begin to arise in the performed experiments 
  so that  
  any improvements in the experiment, typically with 
  an amplification greater than 
  100, 
  should require the 
  nonlinear theory in translating the observed value to 
  the original displacement. 
\end{abstract}


\maketitle

\sect{Introduction}
The standard theory of measurement in quantum mechanics deals with the
situation that one performes a measurement on a quantum state 
to obtain a measured value and the resulting state according to
certain probabilitic laws. 
It was established by von Neumann~\cite{vonneu} 
in the case of projective 
measurements 
and generalized later to non-projective
measurements~\cite{kraus,davies,ozawa}. 
In experiments as well as in theory, 
weak measurements, where the system is weakly coupled 
with, hence weakly disturbed by, the measuring device, 
have been widely considered and have proved to be useful. 

Aharonov, Albert, and Vaidman (AAV)~\cite{aharo} 
proposed a particular type of weak measurement
which is characterized by 
the pre- and post-selection (PPS) of the system. 
One prepares the initial state $\ket i$ of 
the system 
and that $\ket {\Phi_i}$ of the device, 
and after a certain interaction between the system and the meter, 
one post-selects a state $\ket f$ of the system 
and reads the meter value. 
If one measures an observable $A$ of the system, 
one obtains the {\em weak value}\/ 
\begin{align}
  A_w:=\f{\bra{f}A\ket{i}}{\bkt fi}. 
  \label{eq-wv-def}
\end{align}
A peculiar feature of the weak value is that 
it can take 
``strange'' values 
which are outside the range of the eigenvalues of $A$ 
and may even be complex. 
One can easily see from 
\Ref{eq-wv-def} that when $\ket i$ and $\ket f$ are almost
orthogonal, the absolute value of the weak value can be 
arbitrarily large. 
This also means that when the weak value $A_w$ 
becomes very large, 
the probability $|\bkt fi|^2$ 
of successfull post-selection of the state
$\ket f$ becomes very small. 
For example, Aharonov et al.~\cite{aharo} has theoretically 
demonstrated 
that a spin component of a spin-1/2 particle 
can have a weak value of 100. 
Related phenomena have been confirmed in quantum optics~\cite{opt}. 
There are attempts to understand the meaning of $A_w$ 
by formulating a general theory~\cite{jozsa,shi} 
and by examining 
negative probability and 
strange weak values~\cite{aharo-hardy,wil,hos}. 

From these features, we 
may 
say that {\em the weak measurement allows 
one to trade the probability of 
succesful post-selection for 
the magnitude of 
a physical quantity.} 
Recently, 
the weak measurement were applied to precision measurements, 
which 
took advantages of the ``trade'' above. 
Hosten and Kwiat~\cite{hosten} 
measured 
the spin Hall effect of light 
by amplifying 
the small displacement of the light ray by weak measurement. 
Dixon et al.~\cite{dixon} were able to measure 
a small tilt of $400$frad of a mirror 
in a Sagnac interferometer. 
There the observed values are proportional to the weak value 
\Ref{eq-wv-def} of some physical quantity $A$. 
Therefore the amplification
can be arbitrarily large in principle. 
The conventional analysis of the weak measurement, as will be reviewed
shortly, assumes a linear approximation with respect to the coupling
between the system and the meter. Thus it is worth pursuing the true
behavior of the amplification factor in the full analysis. 
Recently, Wu and Li~\cite{wu} extended the general formulation by
Jozsa~\cite{jozsa} and 
gave a formal power expansion analysis of the weak
measurement. They also discussed its effect with a simple example. 

In this paper, we perform a direct analysis of the amplification by
the weak measurement on the Sagnac interferometer, 
up to all orders of the coupling between the system and the meter. 
In particular, 
it is shown that the measured value as well as the amplification
factor  
are bounded in the limit that the PPS states become orthogonal. 
One therefore 
{\em cannot}\/ trade the probability of successful post-selection
for the amplification as much as one wants. 
We also show that in the performed experiments the nonlinear behavior
starts to appear so that improvements of the amplifier 
will require the nonlinear analysis. 

\begin{figure}[tbp]
 \centering
  \includegraphics[width=.41\columnwidth]%
  {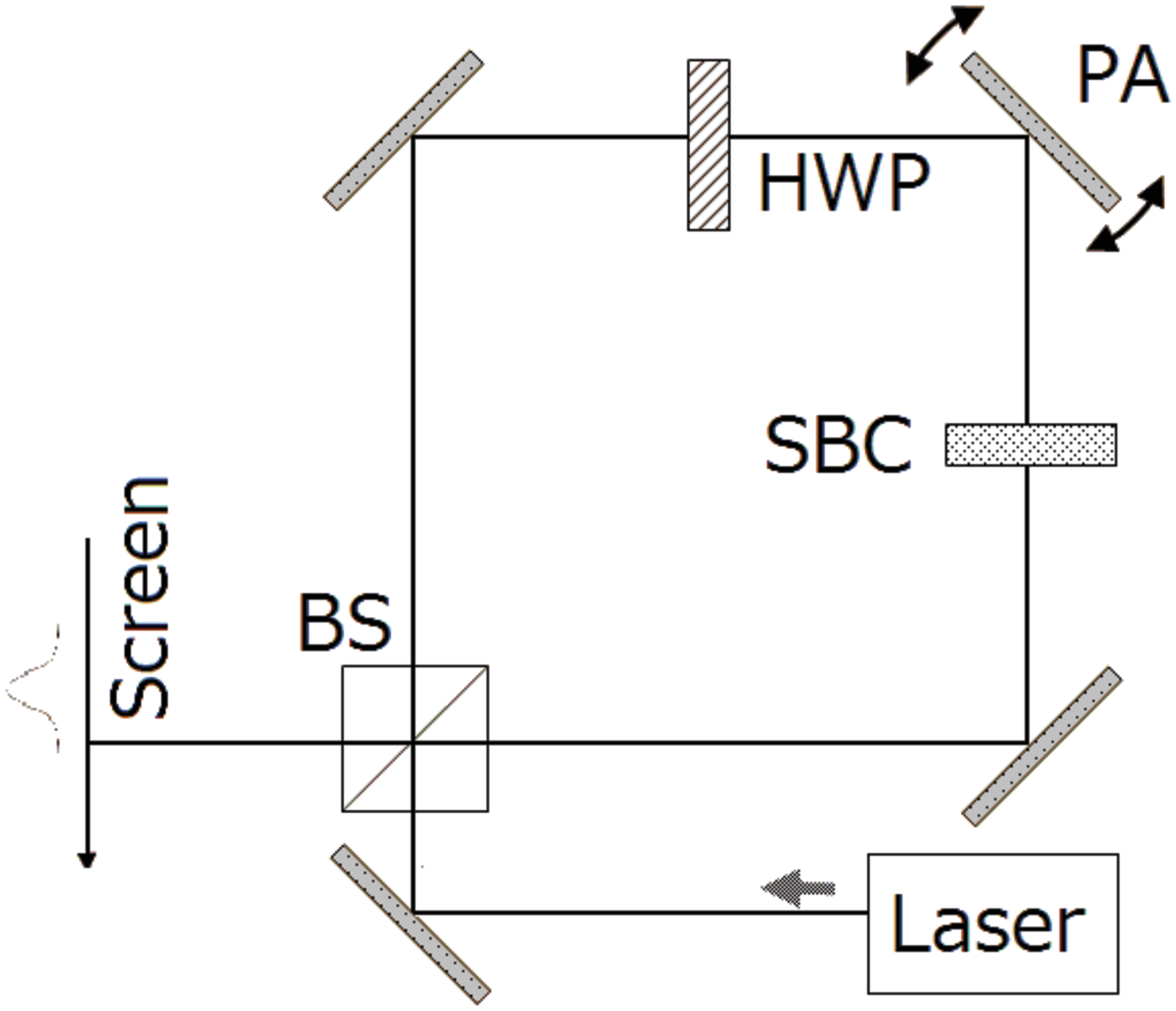}
  \caption{
    Amplifier by 
    the Sagnac interferometer. 
    The input laser beam is divided 
    by the beam splitter (BS) and come back 
    again 
    after given a transverse momentum and 
    a phase difference. 
    The momentum is given by a tilted mirror 
    controlled by a piezo-actuator (PA). 
    The phase difference is given 
    by 
    the 
    SBC 
    under a 
    polarization control. 
    One measures the beam position at the ``screen,'' 
    which is usually a CCD detector. 
  }
  \label{fig-sagnac}
\end{figure}

\sect{Amplification by Sagnac interferometer}
The precision measurement by 
Sagnac interferometer~\cite{dixon}
makes use of the weak measurement where 
the which-path 
degree of freedom is 
the measured system 
and the transverse displacement of the laser is 
the measuring device (meter). 
The input laser is devided by the beam splitter 
to the clockwise path $\ket\cw$ and the anticlockwise one
$\ket\ccw$ (Figure~\ref{fig-sagnac}).  
The beams go through the beam splitter again and all photons 
go out to the port where the input beam comes in (bright port). 
The other port is called the dark port. 
If there is a phase difference between the clockwise and anticlockwise
beams, there are photons coming to the dark port. 
This is done in the following manner. 
First, the input beam is polarized horizontally. 
Second, 
the polarization is changed to vertical by the half-wave plate 
either before or after the beam passes 
the Soleil-Babinet compensator (SBC) 
depending on which path 
the photon takes. 
Third, the SBC 
gives a phase difference $\phi$ 
between the horizontally and vertically polarized beams. 
Finally, one measures the transverse displacement in the dark port 
and obtain its expectation value $\av x$. 
This method gives an amplification of the small tilt of the
mirror. 

The initial state of the 
which-path 
degrees of freedom is given by 
$
  \ket{i}=
  \f1{\sqrt2}(e^{-i\phi/2}\ket{\cw}+ie^{i\phi/2}\ket{\ccw}). 
$ 
When $\phi=0$, the dark port is completely dark. 
The initial state for the transverse dispacement is 
$
  \ket{\Phii}=\int \d x\ket{x}\Phii(x). 
$
Thus the initial state of the whole system is given by 
\begin{align}
  \ket{\Psii}
  =\ket{\Phii}\ket{i}. 
\end{align}
The tilt of the mirror gives a transverse momentum $\pm k$, 
where the signature depends on the path of the beam, 
so that 
the effect is described by a unitary operator $e^{-ikxA}$, 
where $A:=\kb\cw-\kb\ccw$ is the operator which distinguishes the 
paths of the beam.

The post-selected state is the dark port state
$
  \ket{f}=
  \f1{\sqrt2}(i\ket{\cw}+\ket{\ccw}). 
$
Then the final state of the meter is given by  
$
  \ket{\Phif}
  =\bkt {f}{\Psi}
  =
  \int \d x\ket{ x}\Phii( x)\bra{\psif}e^{-ik x A}\ket{\psii}. 
$
The standard argument of the weak measurement is to 
expand $\bra{\psif}e^{-ik x A}\ket{\psii}$ and write $\bra fA\ket i$ 
by the weak value $A_w$ and regroup those to an exponential function, 
$
  \bra{\psif}e^{-ik x A}\ket{\psii}
  \simeq
  \bkt fi (1-ikx A_w)
  \simeq
  \bkt{\psif}{\psii}e^{-ikx A_w}. 
$
One thereby has 
\begin{align}
  \ket{\Phif}
  \simeq
  \bkt fi
  \int \d x\ket{ x}\Phii( x)e^{-ik x A_w}. 
\end{align}
This gives the final read of the meter, 
\begin{align}
  \av x
  =\f{\bra\Phif x\ket \Phif}{\bkt\Phif\Phif}
  \simeq
  2ka^2\Im A_w
  =
  2ka^2\cot\f\phi2, 
  \label{eq-avx-lin}
\end{align}
where 
the wave funtion $\Phii(x)$ is assumed to be even, 
and $A_w=i\cot\f\phi2$ 
has 
been used. 
The length $a=\sqrt{\av {x^2}_i}$ 
is the input beam radius, 
where $\av \bullet_i$ denotes the
expected value with respect to the initial state. 
In the case of diverging paraxial beam performed in the
experiments \cite{dixon}, 
the result \Ref{eq-avx-lin} is modified 
as
\begin{align}
  \av x=2k\sigma[(1-\gamma) a+\gamma \sigma]\Im A_w, 
  \label{eq-avx-mod}
\end{align}
where $\gamma:=\llm/(\llm+\lmd)$ 
with $\llm$ and $\lmd$ being the path lengths 
between the lens and the mirror, 
and between the mirror and the detector, 
respectively, 
and $\sigma$ is the 
radius 
of the beam at the detector. 
Note that \Ref{eq-avx-mod} is consistent with \cite{jozsa}; 
in particular, the imaginary part of a weak value gives a shift in $x$
when the coupling Hamiltonian is proportional to $xA$. 

Dixon et al. defined the amplification factor 
\begin{align}
\AA=\f{|\av x|}{\delta}, 
\label{eq-amp-fac-def}
\end{align}
as the ratio 
of the displacement $\av x$ compared with that $\delta $ 
without interferometer, 
where $\delta=k\lmd/k_0$ with $\lmd$ being the length of the path from
the mirror to the detector and $k_0$ being the wavelength of the
laser. 
Eq. \Ref{eq-avx-mod} shows that the displacement $\av x$ as well as
the 
amplification factor $\AA$ are proportional to the weak value 
\Ref{eq-wv-def}. 
This implies that those quantities diverges in the limit $\phi\to0$ 
when the initial and final states become orthogonal. 
Though in this limit the output power, which is proportional to 
$|\bkt fi|^2=\sin^2\f\phi2$, become infinitesimally small 
and it becomes extremely difficult to perform an
experiment, but apart from that, 
the displacement $\av x$ and the amplification $\AA$
can in principle become arbitrarily large. 

In the rest of the paper, 
we shall calculate 
how the measurement result in the weak 
measurement differs from the weak value and show that 
$\av x$ and $\AA$ actually do not diverge even in the limit 
$\bkt fi\to0$.

\sect{Nonlinear effect}
Let us derive the amplified displacement $\av x$ and the amplification
factor $\AA$ in all orders of $k$. 
Let $\xl$, $\xm$ and $x$ be the transverse 
displacement at the lens, mirror and detector, 
respectively. 
We have 
$
  {\xl}/{\si}=
  {\xm}/{\llim}=
  {x}/{\lid}, 
$
where 
$\si$ be the image distance, 
and $\llim$ and $\lid$ are the path lengths between 
the image and the mirror, and between 
the image and the detector, respectively. 
Let the initial variances of 
$\xl$, 
$\xm$ and 
$x$
be 
$a^2$, $a_m^2$ and $\sigma^2$. 
Then we have $\xl/a=\xm/\am=x/\sigma$ 
and
$
\am=(1-\gamma) a+\gamma\sigma. 
$
It 
is the most 
convenient to use $\xm$ in the calculation 
because the nontrivial state evolution is caused by the tilted mirror. 

The initial state is given by 
\begin{align}
  \ket{\Psii}
  =
  \ket{\Phii}
  \ket{\psii}, 
  \quad
  \ket{\Phii}=\int \d\xm\ket{\xm}\Phimi(\xm), 
\end{align}
where $\Phimi(\xm)$ is the initial wavefunction at the mirror. 
[The relation to the initial wavefuntion 
$\Phili(\xl)$ at the lens 
is given by 
$\Phimi(\xm)=\Phili(\xl)=\Phili\paren{{\si}\xm/{\llim}}$.] 
The photons gain a transverse momentum at the mirror, which is
described by a unitary operator $e^{-ik\xm A}$. 
The state of the system and the meter becomes 
\begin{align}
  \ket{\Psi}=
  \int \d\xm\ket{\xm}\Phimi(\xm)e^{-ik\xm A}\ket\psii. 
\end{align}
With the post-selected state 
$\ket\psif$ of the system, 
the resulting state of the meter reads
\begin{align}
  \ket{\Phif}
  =\bkt {\psif}{\Psi}
  =
  \int \d\xm\ket{\xm}\Phimi(\xm)\bra{\psif}e^{-ik\xm A}\ket{\psii}, 
  \label{eq-Phif}
\end{align}
where the state is not normalized. 
From $A^2=1$, we can exactly calculate 
$\bra{\psif}e^{-ik\xm A}\ket{\psii}$ 
to all orders: 
\begin{align}
  &
  \bra{\psif}e^{-ik\xm A}\ket{\psii}
  \nn
  &=
  \sum_{n=0}^\infty\f{(-ik\xm)^{2n}}{(2n)!}
  \bkt{\psif}{\psii}
  +
  \sum_{n=0}^\infty\f{(-ik\xm)^{2n+1}}{(2n+1)!}
  \bra{\psif}A\ket{\psii}
  \nn
  &=
  \bkt{\psif}{\psii}\paren{\cos k\xm-iA_w\sin k\xm}. 
  \label{eq-fi}
\end{align}
From \Ref{eq-Phif} and \Ref{eq-fi}, 
we have 
\begin{align}
  \av{\xm^n}
  &=
  {}\f{\bra{\Phif}\xm^n \ket{\Phif}}
  {\bkt{\Phif}{\Phif}}
  =
  \f
  {
  \av{\xm^n |\cos k\xm-iA_w\sin k\xm|^2}_i
  }
  {
  \av{ |\cos k\xm-iA_w\sin k\xm|^2}_i
  }. 
\end{align}
Thus we obtain a formula for 
the expectation value of any moment of $x$, 
\begin{align}
  &\av {x^n}=\paren{\f{\sigma}{\am}}^n\av{\xm^n}
  =\paren{\f{\sigma}{\am}}^n
   \nn
  &\cdot
  \,
  \f
  {
    \a_+\av {\xm^n}_i+
    \a_- \av{\xm^n\cos2 k\xm}_i
    +
    \Im A_w\av{\xm^n \sin 2k\xm}_i
  }
  {
    \a_+ +
    \a_-\av{\cos2 k\xm}_i
    +
    \Im A_w \av{\sin 2k\xm}_i
  }, 
  \label{eq-avx}
\end{align}
where $\a_\pm:=(1\pm|A_w|^2)/2$. 
The expectation value $\av x$ is given by \Ref{eq-avx} with $n=1$. 
Then the amplification factor is given by \Ref{eq-amp-fac-def}.
The amplification 
factor depends on $k$, namely, the amplification is nonlinear. 
We note that
we were able to 
write the result \Ref{eq-avx} in terms only of $A_w$, 
by virtue of $A^2=1$; 
otherwise the result would depend on $\bra f{A^k}\ket i$. 

In the present model, 
the weak value is pure imaginary, 
$A_w=i\cot\f\phi2$. 
For simplicity, we assume that 
$|\Phimi(\xm)|^2$ is an even function. 
Then 
from \Ref{eq-avx}, 
we have a simple expression for the expectation value: 
\begin{align}
  \av x=
  \f{\sigma}{\am}
  \,
  \f
  {
    \sin\phi\av{\xm \sin 2k\xm}_i
  }
  {
    1 
    -\cos\phi\av{\cos2 k\xm}_i
  }. 
  \label{eq-avx-even}
\end{align}
This formula gives the measured value obtained from 
the weak measurment, 
for arbitrary 
$\Phimi(\xm)$. 
In the lowest order in $k$, 
\Ref{eq-avx-even} yields
$
\av x=
2k\am\sigma \cot\f\phi2, 
$
which reproduces the result \Ref{eq-avx-mod} by Dixon et al.
When $\phi$ varies, one can see that 
$|\av x|$ attains its maximum 
\begin{align}
  |\av x|_{\max}
  =
  \f{\sigma}{\am}
  \f{
    \av{\xm \sin 2k\xm}_i
  }
  {\sqrt{1-\av{\cos2 k\xm}_i^2} }, 
  \label{eq-avxmax}
\end{align}
at 
$\cos\phi=\av{\cos2 k\xm}_i$. 
In particular, 
when $k\am$ is small (recall that $\av {\xm^2}=\am^2$), 
\Ref{eq-avxmax} implies 
\begin{align}
  |\av x|_{\max}
  =\sigma
  +O(k\am)^2. 
  \label{eq-avxmax-app}
\end{align}

One can see from \Ref{eq-avx-even} 
and \Ref{eq-avxmax} 
that both the displacement $|\av x|$
and the amplification factor $\AA$ 
are {\em bounded}\/ even when 
the 
PPS states become orthogonal. 
In the linear analysis of 
the weak measurement, the 
obtained value 
was proportional to the weak value. 
This implies that 
it 
diverges when $\phi\to0$, 
so that one can achieve arbitrarily 
large amplification factor $\AA$, 
though the power of the output beam becomes extremely small. 
The full order result shows that the 
measured value of the weak measurement 
is not proportional to the weak value. 
The observed value $\av x$ does not diverge but go to zero 
when $\phi\to0$. 
Thus, when the PPS states become orthogonal, 
the amplification factor $\AA$ is bounded (in fact, vanishes). 
In other words, 
one cannot achieve infinite amplification 
even one compromises on the power of the output beam or 
the probability of successful post-selection.

\begin{figure}[t]
  \centering
  \includegraphics[width=.75\columnwidth]%
  {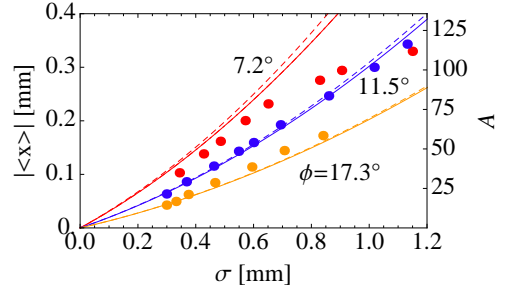}
  \caption{(Color online) The 
    meter value $|\av x|$ and 
    the amplification factor $\AA$ 
    in the weak measurement 
    as a function of the beam radius $\sigma$ at the detector, 
    for three values of $\phi$. 
    The points are the experimental data from \cite{dixon}. 
    The solid curve is the 
    theoretical meter value \Ref{eq-avx-even} by
    the nonlinear analysis. 
    The dashed curve is the one \Ref{eq-avx-mod} by the
    linear analysis which is 
    proportional to the weak value $A_w$. 
  }
 \label{fig-meter-val-vs-sigma}
\end{figure}

\begin{figure}[t]
  \centering
  \includegraphics[width=.57\columnwidth]%
  {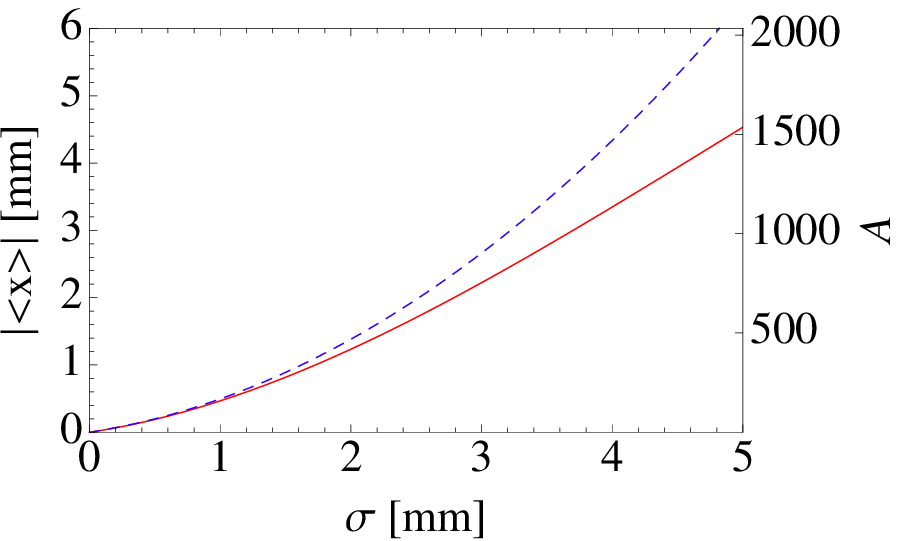}
  \caption{(Color online) The same as the case
    $\phi=7.2\degr$ in 
    Figure~\ref{fig-meter-val-vs-sigma} but with larger scale. 
    The solid (dashed) curve is $|\av x|$ by nonlinear (linear) 
    analysis 
    \Ref{eq-avx-even} [\Ref{eq-avx-mod}]. 
  }
 \label{fig-meter-val-vs-sigma-largerscale}
\end{figure}

To evaluate \Ref{eq-avx-even} 
or \Ref{eq-avxmax} 
explicitly, 
let us consider the case that the beam is Gaussian, 
\begin{align}
  \di|\Phimi(\xm)|^2=\f{e^{-x^2/2\am^2}}{\sqrt{2\pi}\, \am}. 
\end{align}
We have 
$\av{\cos2 k\xm}=e^{-2(k\am)^2}$ and
$\av{\xm\sin2 k\xm}=2k\am^2e^{-2(k\am)^2}$ 
so that 
\begin{align}
  &\av x
  =
  \f{2k\am\sigma\sin\phi}{e^{2(k\am)^2}-\cos\phi}, 
  \\
  &|\av x|_{\max}
  =
  \f{2k\am\sigma }
  {\sqrt{e^{4(k\am)^2}-1 }}
  \ \text{  at } \cos\phi=e^{-2(k\am)^2}. 
\label{eq-avxmax-gau}
\end{align}
Recall that $\am=(1-\gamma) a+\gamma\sigma$. 

\begin{figure}[tb]
  \centering
  \includegraphics[width=.63\columnwidth]%
  {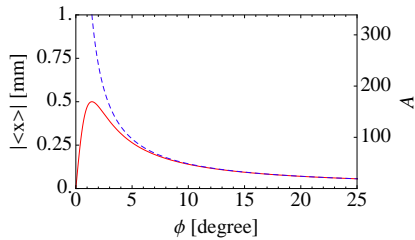}
  \caption{(Color online) The 
    theoretical 
    meter value $|\av x|$ and  the amplification factor $\AA$
    in the weak measurement 
    as a function of the phase difference $\phi$. 
    The 
    solid curve is from 
    the nonlinear analysis 
    \Ref{eq-avx-even}.  
    The dashed curve is 
    from the
    linear analysis 
    \Ref{eq-avx-mod} which is 
    proportional to the weak value $A_w$. 
  }
 \label{fig-meter-val-vs-phi}
\end{figure}

Figure~\ref{fig-meter-val-vs-sigma} 
shows the theoretically derived meter value
$|\av x|$ in the weak measurement as a function of the beam radius
$\sigma$ at the detector. 
The parameters are chosen the same as Figure~2 of 
the experiment carried out in \cite{dixon}, 
$a=640\micm$, $\gamma=.296$, $k=20.8\m^{-1}$. 
For the larger phase differeces $\phi$, the value $|\av x|$ 
in the nonlinear analysis \Ref{eq-avx-even} 
essentially has no difference 
to that in the linear analysis \Ref{eq-avx-mod}. 
For the smallest phase difference $\phi$, 
the nonlinear value gives slightly better fit to the experimental
data but  still does not fit well to them. 
(The authors of \cite{dixon} suggested that the difference 
may be due to 
the effect of stray light.) 
As a reference, Figure~\ref{fig-meter-val-vs-sigma-largerscale} 
shows the same theoretical meter values in a larger scale 
(though it may not be realistic in the performed experiment). 
It can be seen that, in the setup in \cite{dixon}, 
the difference between linear and nonlinear results 
would be significant when $\sigma$ is larger than 1-2 mm 
when the amplification is larger than several hundred.

Figure~\ref{fig-meter-val-vs-phi} shows the theoretically derived meter value 
$|\av x|$ in the weak measurement, as a function of the phase
difference $\phi$. 
We chose a typical value of $\sigma=500\micm$. 
One can verify that 
when the PPS states become orthogonal, $\phi\to0$, 
the $\av x$ in the linear analysis \Ref{eq-avx-mod}
diverges while 
that in the nonlinear analysis \Ref{eq-avx-even}
converges to zero. 
One can also verify $|\av x|_{\max}\lesssim \sigma$ 
implied by \Ref{eq-avxmax-app} or \Ref{eq-avxmax-gau}. 
One finds that the smallest phase difference $\phi=7.2^\circ$ 
chosen in the experiment in \cite{dixon} 
is on the border of valid linear analysis, beyond which one needs 
nonlinear analysis to measure the original tilt angle of 
the mirror.


\sect{Conclusion}
We have performed an all-order analysis for the amplifier by Sagnac
interferometer. 
In the linear analysis, 
the measured meter value $\av x$ was proportional to the weak
value \Ref{eq-wv-def} and hence diverges in the limit $\phi\to0$ when 
the pre- and post-selected states be orthogonal. 
In the nonlinear analysis, the measured value 
is no longer proportional to the weak 
value and vanishes in the limit $\phi\to0$. 
It is suggested that nonlinear analysis of the weak measurement is
necessary if we improve the amplifier by the Sagnac interferometer. 
In a realistic setting, we have seen that this 
happens when the amplification factor $\AA$ 
goes beyond a typical value of roughly 
100, 
and that the maximum possible value of $\AA$ resulting from 
the nonlinear analysis is several hundred.

We would like to express our sincere thanks 
to Professor Akio Hosoya for valuable suggestions.

\end{document}